\documentclass[proceedings]{JHEP3}

\conference{International Europhysics Conference on HEP}                

\title{NONCOMMUTATIVITY OF BOUNDARY CLOSED STRING COORDINATES FOR AN OPEN MEMBRANE ON p-BRANE }
\author{Ashok Das$^\dagger$,
\speaker{J. Maharana$^\ddagger$} and  A. Melikyan$^\dagger$ }
\author{{$^\dagger$}                                           
{\it Department of Physics and Astronomy\\University of Rochester, Rochester, NY 14627-0171, USA.}\\
      E-mail: \email{das@pas.rochester.edu},
      \email{arsen@pas.rochester.edu}}
\author{{$^\ddagger$}
\it {Theory Division, KEK, Tsukuba-shi, Japan.}\\
      E-mail: \email{maharana@post.kek.edu}}

\abstract{We study the dynamics of an open membrane with a
cylindrical topology, in the background of a constant  three form.
We use the action, due to Bergshoeff, London and Townsend, to
study the noncommutativity properties of the boundary string
coordinates. The constrained Hamiltonian formalism due to Dirac is
used to derive the  noncommutativity of coordinates. The chain of
constraints is found to be finite for a suitable gauge choice.}

\begin{document}

 \section{Introduction:}

Recently, the study of noncommutative geometry, from the
perspective of string theory, has attracted considerable
attention. The noncommutativity of the target space coordinates
becomes manifest when a constant background NS two form potential
is introduced along the D-brane \cite{noncom1}. In the presence of
the two-form potential, the end points of the open strings
attached to the D-brane do not commute. It is natural, therefore,
to examine the corresponding issue when an open membrane ends on a
D-brane under an analogous situation. In this talk, we
present the noncommutativity property of an open membrane-brane
configuration from a different perspective; details of the result
are given in \cite{dmm}. We adopt a modified version of the
action due to Bergshoeff, London and Townsend \cite{blt}. With the
modified form of the action, we are able to make some head way
with the computation of the matrix of constraints as well as the
evaluation of the Dirac brackets (DB) in a systematic manner,
without linearizing the action as was done at the outset in
\cite{membr2}. Surprisingly, however, we find that with an
alternate, suitable choice of gauge, the constraint chain for the
same membrane system terminates. In other words, in this alternate
gauge,  after a finite number of iterations, new constraints are
not generated.

\section{The Action}

The action for a membrane, interacting with an anti-symmetric
background field $C_{MNP}$ can be described by  a Nambu-Goto
action
\begin{equation}
S=T\,\int_{\Sigma_{3}}d^{3}\xi\left(\sqrt{g}-\frac{1}{6}\varepsilon
^{ijk}\partial_{i}X^{M}\partial_{j}X^{N}\partial_{k}X^{P}A_{MNP}\right)
\label{4}
\end{equation}
where  $A=C+dB$. Here $M,N,P=0,1,..,10$ are indices of the
$11$-dimensional target space, $\xi=(\tau,\sigma_{1},\sigma_{2})$
are the coordinates of the world volume of the membrane with the
corresponding indices taking values $i,j,k = 0,1,2$, $g=\det
g_{ij}$, where $g_{ij}= G_{MN}\partial_{i}X^{M}\partial_{j}X^{N};$
is the  induced metric on the membrane. For simplicity, we are
going to choose both $G_{MN}$ and $A_{MNP}$ to be constants (In
fact, we will choose $G_{MN}=\eta_{MN}$ from now on). An alternate
description for the membrane is through the first order action of
the form (due to Bergshoeff, London and Townsend)
\cite{blt},\cite{jmward}
\begin{equation}
S_{BLT}=\int_{\Sigma_{3}}d^{3}\xi\,\frac{1}{2V}(g-\widetilde{\mathcal{F}}^
{2}) \label{5}
\end{equation}
Here we have defined
\begin{equation}
\widetilde{\mathcal{F}}\equiv
\varepsilon^{ijk}\widetilde{\mathcal{F}}_{ijk} =\varepsilon^{ijk}
(F_{ijk}+\frac{1}{6}A_{ijk})
\label{6}%
\end{equation}%
where
\begin{eqnarray}
F_{ijk} & = &\partial_{\lbrack i}U_{jk]} = \partial_{i}U_{jk} +
{\rm
cyclic} \label{7}\\
A_{ijk} & = &
\partial_{i}X^{M}\partial_{j}X^{N}\partial_{k}X^{P}A_{MNP}\label{8}%
\end{eqnarray}
with $A_{MNP}$ defined earlier. Clearly, $V(\xi)$ and
$U_{ij}(\xi)$ are auxiliary field variables. It can be shown that
the dynamical equation for the coordinates and boundary conditions
coincide with the ones following from the action in (\ref{4}).

Due to the cylindrical topology of the membrane, in the presence
of p-branes, the boundary condition reduces to
\begin{equation}
\sqrt{g}g^{1j}\partial_{j}X_{\mu}-\frac{1}{2}\varepsilon^{1jk}
\partial_{j}X^{\nu}\partial_{k}X^{\rho}A_{\mu\nu\rho}
\mid_{\sigma_{1}=0,\pi}=0,~~~~~~~~~~~~~~ \mu,\nu=0,1,...,p
\label{16}
\end{equation}

\begin{equation}
X^{a}=x_{0}^{a},~~~~~~~~~~~~~ a=p+1,...,10
\label{17}
\end{equation}
where the coordinates, $x_{0}^{a}$, specify the  positions of the
p-branes  at the two boundaries $\sigma_{1}=0,\pi$.
\bigskip

\section{Constraint analysis:}

We take the first order action (\ref{5}) as the starting point of
our canonical description of the system. Let us note that the
Lagrangian density of the action (\ref{5}) is singular. Therefore,
the velocities cannot be expressed in terms of the phase space
variables. We will choose a gauge condition which brings out an
interesting feature of our analysis, namely, that with a  suitable
gauge choice, the chain of constraints can terminate.  Let us look
at the action (\ref{5}) in the gauge
\begin{equation}
g_{0a} = 0,\qquad a=1,2\label{22'}
\end{equation}
In this gauge,  action (\ref{5}) takes the form
\begin{equation}
S = \int_{\Sigma_{3}} d^{3}\xi\,\frac{1}{2V} \left(\overline{g}
\dot{X}^{M}\dot{X}_{M} -
\widetilde{\mathcal{F}}^{2}\right)\label{22''}
\end{equation}
Adding boundary constraints, it is now straightforward to
determine the complete set of primary constraints:
\begin{eqnarray}
\varphi_{1} & = & P_{V}\approx 0\nonumber\\
\varphi_{2}^{a} & = & \Pi^{(U)oa}\approx 0\nonumber\\
\varphi_{3\mu} & = &
(\overline{g}^{a1}\partial_{a}X_{\mu}\mathcal{P}
^{2}+\frac{1}{3}\Pi^{(U)a1}\mathcal{P}^{\nu}\partial_{a}X^{\lambda}A_{\mu\nu\lambda
})\delta(\sigma_{1})\approx 0\nonumber\\
\varphi_{4\mu}  & = &
(\overline{g}^{a1}\partial_{a}X_{\mu}\mathcal{P}
^{2}+\frac{1}{3}\Pi^{(U)a1}\mathcal{P}^{\nu}\partial_{a}X^{\lambda}A_{\mu\nu\lambda
})\delta(\sigma_{1}-\pi)\approx 0 \label{28}
\end{eqnarray}
One can obtain the Hamiltonian for the system to be
\begin{equation}
H=\frac{V}{2\overline{g}}\mathcal{P}^{2}-\frac{V}{72}(\Pi^{(U)ab}
\varepsilon_{ab})^{2}+2\Pi^{(U)ab}\partial_{a}U_{0b}+c\varphi_{1}+
k_{a}\varphi_{2}^{a}+\lambda^{\mu}\varphi_{3\mu}+\widetilde{\lambda}^{\mu
}\varphi_{4\mu} \label{29}
\end{equation}
where $c,k_{a},\lambda^{\mu},$ and $\widetilde{\lambda}^{\mu}$ are
Lagrange multipliers.

The analysis for the consistency of constraints can now be carried
out in a straightforward manner. The consistency conditions lead
to
\begin{eqnarray}
\lambda^{\mu} & = & \widetilde{\lambda}^{\mu} = 0\nonumber\\
\varphi_{5} & = & \frac{{\mathcal{P}}^{2}}{2\overline{g}} -
\frac{1}{72} \left(\Pi^{(U)ab}\varepsilon_{ab}\right)^{2} \approx
0\label{30}
\end{eqnarray}
\begin{equation}
\varphi^{a}_{6}  =  \partial_{b}\Pi^{(U)ba} \approx 0 \label{32}
\end{equation}
This is the analog of Gauss' law in electrodynamics  and it can be
easily checked that the consistency of these constraints leads to
no new constraints. Consistency of the boundary constraint
$\varphi_{3\mu}\approx 0$,
\[
\dot{\varphi}_{3\mu}=\left\{\varphi_{3\mu}, \int H\right\} \approx
0
\]
leads to the secondary constraint,
\begin{eqnarray}
\varphi_{7\mu} & = &
\delta(\sigma_{1})\left[\overline{g}^{a1}\partial
_{a}[\frac{V}{\overline{g}}\mathcal{P}_{\mu}]+\varepsilon^{ab}\partial
_{a}X_{\mu}\partial_{(b}X^{\lambda}\partial_{2)}[\frac{V}{\overline{g}
}\mathcal{P}_{\lambda}]+\frac{V}{3\overline{g}}\Pi^{(U)a1}A_{\mu\nu\lambda
}\mathcal{P}^{\nu}\partial_{a}\mathcal{P}^{\lambda}\right.\nonumber\\
 & &
\left.
-\frac{1}{3}\Pi^{(U)a1}A_{\mu\nu\lambda}\partial_{a}X^{\lambda}
\partial
_{c}[V\overline{g}^{bc}\partial_{b}X^{\nu}]\right]\label{36}
\end{eqnarray}

Since the Hamiltonian (\ref{29}) contains a term of the form $c
P_{V}$ and the secondary constraint $\varphi_{7\mu}$ depends on
$V$ as well as $\partial V$, consistency of this new constraint
\[
\dot{\varphi}_{7\mu} = \left\{\varphi_{7\mu}, \int H\right\}
\approx 0
\]
simply determines the Lagrange multiplier $c$ and leads to no
further constraint. An identical analysis goes through for the
constraint $\varphi_{4\mu}$ at the other boundary and generates
only one secondary  constraint $\varphi_{8\mu}$, whose structure
is identical to that of $\varphi_{7\mu}$ except that it is at the
other boundary.

\section{Dirac brackets:}

Since we have determined all the constraints of our theory, it is
now straightforward, in principle, to determine the Dirac brackets
\cite{dirac} . However, we note that the boundary constraints are,
in particular, highly nonlinear and, consequently, evaluation of
the inverse of the matrix of constraints is, in general,  a very
difficult problem. Things, however, do simplify enormously if we
use a weak field approximation for $A_{\mu\nu\lambda}$.The
constraints $\varphi_{3\mu},\varphi_{7\mu}$ are second class and,
therefore, the Dirac bracket between the coordinates takes the
form
\begin{eqnarray}
\{X_{\mu}(\sigma),X_{\nu}(\sigma^{\prime})\}_{D} & = & -\int
d^{2}\sigma'' d^{2}\sigma'''\,\{X_{\mu}(\sigma
),\phi_{A}(\sigma'')\}C^{-1}{}^{AB}(\sigma'',\sigma''')\nonumber\\
 &  & \qquad \times \{\phi_{B}(\sigma'''),X_{\nu}(\sigma')\}
\label{DB1}
\end{eqnarray}
where $\phi_{A}\equiv(\varphi_{3\mu},\varphi_{7\mu})$ and
\begin{equation}
C_{AB}(\sigma,\sigma')=\left(
\begin{array}
[c]{cc}%
\{\varphi_{3\mu}(\sigma),\varphi_{3\nu}(\sigma')\} &
\{\varphi_{3\mu}(\sigma),\varphi_{7\nu}(\sigma')\}\\
\{\varphi_{7\nu}(\sigma'),\varphi_{3\mu}(\sigma)\} &
\{\varphi_{7\mu}(\sigma), \varphi_{7\nu}(\sigma')\}
\end{array}
\right)  \label{C}
\end{equation}

We can, of course, calculate exactly all the brackets entering
into the matrix, $C_{AB}$.  However, determining the inverse
matrix exactly is a technically nontrivial problem. It is here
that the weak field approximation is of immense help (We want to
emphasize that there is no other approximation used in our
derivations.). It is shown  (\cite{dmm}) that in this
approximation the Dirac bracket has the form
\begin{eqnarray}
\{X_{\mu}(\sigma),X_{\nu}(\sigma')\}_{D}  & = &
\left[\widetilde{T}_{\mu\lambda}
\left(\left(G+F\right)^{-1}\right)^{\lambda\rho}S_{\rho\nu}\right.\nonumber\\
  &  &   - \widetilde{S}_{\mu\lambda}\left(\left(\left(G+F\right)^{-1}\right)^
{\lambda\rho}T_{\rho\nu}\right)\label{final}\\
 &  &
+\left.\widetilde{T}_{\mu\lambda}\left(\left((F+G)^{-1}(V+W)(F+G)^{-1}\right)^
{\lambda\rho}T_{\rho\nu}\right)\!\right]\!(\sigma,\sigma')\nonumber
\end{eqnarray}
where the  operators $T,F,G$ are independent on
$A_{\mu\nu\lambda}$, while $S,V,W$ are liner in
$A_{\mu\nu\lambda}$. It is worth emphasizing here that because of
the structure of the boundary constraints the $\sigma_{1}$
coordinate is fixed at the boundary (to be $0,\pi$) and,
therefore, the Dirac bracket, evaluated at equal $\tau$,
effectively depends only on the world volume coordinates
$\tau,\sigma_{2}$. This shows that the boundary string coordinates
indeed become noncommutative  in the presence of an anti-symmetric
background field and what is even more interesting is that they
have a structure that is quite analogous to that in the case of
strings.

\section{Summary}

We have studied an open membrane, with cylindrical geometry,
ending on p-branes. The boundary of the open membrane on the brane
is a closed string. We have adopted a modified action which has
some distinct advantages as discussed in the text. We have treated
the boundary conditions as primary constraints  and have shown
that, one can carry out the Dirac formalism without restricting to
the linearized approximation of the action. We have also
introduced a gauge choice, different from the one adopted in ref
\cite{membr2}, and have shown that the Dirac procedure, in this
gauge, leads to a finite number constraints. As a consequence, we
are able to compute the PB brackets of all  the second class
constraints which are necessary for the evaluation of Dirac
brackets.

\section*{Acknowledgments}
AD would like to thank S. Minwalla for a helpful discussion. JM
would like to acknowledge supports from the Albert Einstein
Institute and warm hospitality of Prof. H. Nicolai where most of
the work was done. He expresses thanks to KEK and Prof. Y.
Kitazawa for the gracious hospitality. This work is supported in
part by US DOE Grant No. DE-FG 02-91ER40685.

\end{document}